\documentclass[
  ,numberedheadings 
  ]
  {aipproc}

\layoutstyle{8x11single}

\newcommand{\vect}[1]{\textbf {\textit{#1}}}

\newcommand{\im}{\mathrm{Im}}
\newcommand{\re}{\mathrm{Re}}

\newcommand{\Slash}[1]{\ooalign{\hfil/\hfil\crcr$#1$}}

\newcommand{\vp}{\vect{p}}
\newcommand{\vk}{\vect{k}}

\newcommand{\comment}[1]{}

\newcommand{\cp}{g}

\newcommand{\mb}{m_b}
\newcommand{\mf}{m_f}
\newcommand{\zetaf}{\zeta_f}
\newcommand{\zetab}{\zeta_b}
\newcommand{\damping}{\zeta}

\begin{document}

\title{Fermion Spectrum at Ultrasoft Region in a Hot QED/QCD Plasma}
\classification{12.20.Ds, 12.38.Cy, 12.38.Mh}
\keywords      {collective excitation, quark-gluon plasma}

\author{Daisuke Satow}{
  address={Department of Physics, Kyoto University, Sakyo-ku, Kyoto 606-8502, Japan}
}

\author{Yoshimasa Hidaka}{
address={Mathematical Physics Lab., RIKEN Nishina Center, Saitama 351-0198, Japan} 
}

\begin{abstract}
  An spectrum of an electron (quark) with ultrasoft momentum ($p\sim\cp^2T$) is discussed at extremely high temperature in quantum electrodynamics (chromodynamics) with a resummed perturbation theory.
We establish the existence of the pole suggested in the previous work of the electron (quark) propagator at ultrasoft region, and find that its pole position is $p_0=\mp |\vp|/3-i\damping$, where $\damping$ is the damping rate which is of order $\cp^2T\ln 1/\cp$, and the expression of its residue.
We also discuss whether the resummation scheme preserves the gauge symmetry.
\end{abstract}

\maketitle


\section{Introduction}
It is considered that the quark-gluon plasma (QGP) is realized in the heavy ion collision experiment and early universe.
Clarifying what kinds of collective excitations exist in the QGP phase is very important to understand the collective character of QGP.
However, as will be explained, not all of the energy scale are well investigated even in the weak coupling region ($\cp\ll 1$).
Therefore unknown collective excitation may exist there.

Here we briefly review the present status of the analysis on elementary and collective excitations.
Notice that the following results of the perturbative analysis are valid not only in quantum chromodynamics (QCD), but also in general fermion-boson system including Yukawa theory, quantum electrodynamics (QED).
At extreme high temperature ($T$) in which the particle masses are negligible, the following collective excitations are expected to exist in each energy region  when the coupling constant is small ($\cp\ll 1$) enough to apply the perturbation theory.
The particle whose energy is of the order of $T$ can be described as almost free particle.
When the energy is of the order of $\cp T$, the characters of the particle such as dispersion relation is very different from that in the vacuum.
Thus they are no longer regarded as single particle excitations and described as collective excitations.
The bosonic excitation is called plasmon, while the fermionic one is called normal fermion \cite{HTL}. (We note that the perturbative expansion used when the energy is $\sim\cp T$ is called Hard Thermal Loop (HTL) approximation.)
Furthermore, a fermion-hole like excitation, which is called plasmino and does not have counterpart in vacuum, is suggested to exist \cite{HTL}.
When the energy is much smaller ($\ll\cp^2T$ or $\sim \cp^2T$), there are hydro modes, which is bosonic, and whose existence is guaranteed by the conservation laws of the fermion number, energy and momentum.

By contrast, it is not well investigated whether any fermionic modes exist in that energy region\footnote{Nevertheless, when the boson has mass which is comparable with $T$, a new fermionic collective excitation was suggested to exist \cite{kitazawa}.
Furthermore, also in the massless case, such mode was suggested in the pioneering paper \cite{lebedev}, though the analysis is inconclusive: the dispersion relation, decay width, and residue were not obtained.}.
This is because an analysis with an ordinary perturbation theory such as HTL approximation is unreliable in this energy region because of an infrared divergence called pinch singularity~\cite{pinch}. 
We analyze the fermion spectrum in this energy region with the method introduced later, which regularizes the pinch singularity, and discuss the existence of new fermionic excitation and its character.

\section{Resummation}
In the following calculation, we adopt the temporal gauge and use the Keldysh formalism \cite{keldysh}.
Only calculation in QCD will be shown; results in QED will be obtained if one does the following replacement in Eq.~(\ref{eq:dispersion-residue}): $Z= e^2/(144\pi^2)$.
The similar analysis in the Yukawa theory is performed in \cite{yukawa}.

In the present case, the pinch singularity appears when the dispersion relations of the boson and the fermion are the same or the decay widths of both particles are zero.
Because we are considering so high temperature case that the boson and fermion masses are negligible, the pinch singularity appears in naive perturbation expansion.
In order to regularize this singularity, we reorganize the perturbation expansion in the following way~\cite{lebedev}: 
Generally, at finite temperature, the dispersion relations of particles are different from those in vacuum.
In the present case, especially, the dispersion relations of the fermion and the boson are different.
Furthermore, they acquire finite decay widths.
By considering these facts, the pinch singularity is expected to be regularized.
If we remember that the dominant contribution to the self-energy comes from the case that the internal momenta is of the order of  $T$, we can obtain the expressions of the thermal masses;
by using that the energy is of the order of $T$, we use the following expressions of the thermal mass of the boson (fermion): $\mb\equiv \cp T\sqrt{N/3+N_f/6}$ ($\mf\equiv \cp T\sqrt{C_f}/2$), where $N$ ($N_f$) is the color (flavor) number and $C_f\equiv(N^2-1)/(2N)$.
The expressions of the decay widths of the fermion ($\zeta_f$) and the boson ($\zeta_b$), which are of the order of $\cp^2T\ln(1/\cp)$ \cite{damping},  have been obtained only at leading log accuracy.
Thus when the bare propagators yield pinch singularity, we replace the bare ones with the following resummed retarded propagators, which contain the information of the thermal masses and the decay widths:
\begin{eqnarray}
G^R_{ij}(k)=G^R(k)\delta_{ij}= -\frac{\delta_{ij}\Slash{k}}{k^2-\mf^2+2ik^0\zeta_f},~~D^{ R \mu\nu}_{ab}(k)=D^{R \mu\nu}(k)\delta_{ab}=-\frac{\delta_{ab}P^{T \mu\nu}(k)}{k^2-\mb^2+2ik^0\zeta_b} .
\end{eqnarray}
Here, the projection operator into the transverse component is defined as follows: $P^{\mu\nu}_T(k)\equiv g^{\mu i}g^{\nu j}(\delta^{ij}-\hat{k}^i\hat{k}^j)$ with $\hat{k}^i\equiv k^i/|\vk|$.
In determining the propagators, we also used the fact that because we are considering the case that the momentum is of the order of  $T$, the longitudinal component of the boson propagator is negligible, and the residues of both propagators are almost same as those in vacuum.
By using this resummed propagator, the pinch singularity is regularized and thus the contribution from the one-loop diagram becomes finite \cite{lebedev,yukawa}.

However, it is not sufficient to include only the one-loop diagram in gauge theory unlike the Yukawa theory \cite{yukawa}.
Since the contribution from any multi-loop diagrams which are ladder-type is the same order as the one-loop diagram, we have to sum up the contribution from all of the ladder diagrams.
To sum up all of the ladder diagrams, we introduce the fermion-boson vertex function $\cp\Gamma^{\mu }(p;k)t^a_{ij}$ defined by the following self-consistent equation:
\begin{eqnarray}
\label{eq:vertex}
\nonumber
\Gamma^\mu(p;k)=\hspace{-8mm}&\gamma^\mu+i\cp^2\int\frac{d^4 l}{(2\pi)^4}
\biggl[-\frac{N}{2} 
\frac{g_{\alpha\beta}-((k-l)_\alpha u_\beta+(k-l)_\beta u_\alpha)/(k^0-l^0)+(k-l)_\alpha (k-l)_\beta/(k^0-l^0)^2}{(k-l)^2}\\
&\times(g^{\mu\alpha}(2k-l)_\nu+g^\alpha_\nu(2l-k)^\mu-g^\mu_\nu(l+k)^\alpha)\gamma^\beta\\
\nonumber
&+\left(C_f-\frac{N}{2}\right)\gamma_\nu\frac{\Slash{l}+\Slash{p}+\Slash{k}}{(l+p+k)^2}\gamma^\mu\biggr]
 (G^R(p+l)D^{S\nu\rho}(-l)+G^S(p+l)D^{R\nu\rho}(-l))\Gamma_\rho(p;l).
\end{eqnarray}
Here $u^\mu\equiv (1,\mathbf{0})$, $t^a$ is the generator of SU($N$) in the fundamental representation, $G^S(k)\equiv(1/2-n_F(k^0))(G^R(k)-(G^R(k))^*)$, and $D^S_{\mu\nu}(k)\equiv(1/2+n_B(k^0))(D^R_{\mu\nu}(k)-(D^R_{\mu\nu}(k))^*)$, where $n_F$ ($n_B$) is the Fermi (Bose) distribution function.
This vertex function enables us to write the contribution from all of the ladder diagrams with the following single expression:
\begin{eqnarray}
\label{eq:selfenergy}
\Sigma^R_{il}(p)=\Sigma^R(p)\delta_{il}=i\cp^2t^a_{ij}t^b_{kl}\int\frac{d^{4}k}{(2\pi)^{4}} \gamma_\mu (D^{S \mu\nu}_{ab}(-k)G^R_{jk}(p+k)
+D^{R \mu\nu}_{ab}(-k)G^S_{jk}(p+k))\Gamma_\nu(p;k).
\end{eqnarray}

Here we stop the explanation of the resummation scheme and perform a consistency check by confirming the scheme preserves the gauge symmetry.
Concretely, we check that the Ward-Takahashi (W-T) identity is satisfied.
One can see that the following equation is valid by multiplying $k_\mu$ to Eq.~(\ref{eq:vertex}):
\begin{eqnarray}
k_\mu \Gamma^\mu(p,k)=\Slash{k}+\Sigma^R(p).
\end{eqnarray}
This equation is the W-T identity itself, at the leading order.

\section{Results and Summary}
We calculated $\Sigma^R(p)$ from Eqs.~(\ref{eq:vertex}) and (\ref{eq:selfenergy}) at the leading order analytically, focusing on the momentum region $|p^0+i\zeta|$,~$|\vp|\ll\cp^2T$, and derived the retarded fermion propagator from that quantity.
As a result, we found that the fermion retarded propagator has a pole in the ultrasoft region ($p\sim \cp^2T$).
This means that we established the fermionic excitation in ultrasoft region suggested in \cite{lebedev}.
In the fermion number $\pm 1$ sector, the dispersion relation, decay width, and residue are
\begin{equation}
\label{eq:dispersion-residue}
\re p^0=\mp\frac{|\vp|}{3},~~
\im p^0=-\damping,~~
Z= \frac{\cp^2}{16\pi^2}\left(\frac{4N_f}{3}+\frac{13N}{6}+\frac{1}{2N}\right)^2,
\end{equation}
respectively.
Here we have introduced $\damping\equiv\zetaf+\zetab$.

\begin{figure}
  \includegraphics[height=.2\textheight]{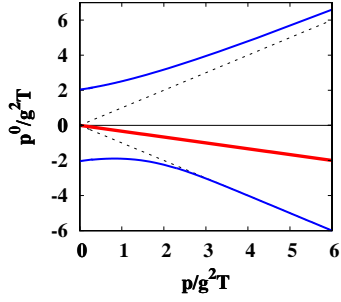}
  \caption{The dispersion relations of the fermionic excitations.
  The blue (solid) lines are the dispersion relations of the normal fermion and the anti-plasmino, while the red (bold) line is that of the ultrasoft fermion mode.
  The dashed lines are the light cone.
  The parameters are set to $\cp=0.2$ and $N=3$.
  Notice that though the plot of the ultrasoft mode is drawn also in the region $|\vp|> \cp^2T$, our analysis is valid only in $|\vp|\ll\cp^2T$.
}
  \label{fig:dispersion}
\end{figure}

From the expression of the dispersion relation, we see that the excitation energy is negative like anti-plasmino \cite{HTL}.
This property suggests that this excitation can be interpreted as anti-fermion hole.
The dispersion relation of that excitation is plotted in Fig.~\ref{fig:dispersion} with those of HTL results.
Notice that the residue given by Eq.~(\ref{eq:dispersion-residue}) is of the order of $\cp^2$.
This is much smaller than the residues of the normal fermion and anti-plasmino, which are of the order of unity \cite{HTL}.

Here we summarize our results.
We analyzed the fermion propagator in the ultrasoft region ($p\sim \cp^2T$) in QED and QCD with a resummed perturbation theory \cite{lebedev,yukawa}.
As a result, a new fermionic excitation was found.
We obtained the expressions of the dispersion relation, decay width, and residue of the new excitation.
We also checked that the derived vertex function, $\Gamma^\mu(p;k)$, and the fermion self-energy, $\Sigma^R(p)$, satisfy the W-T identity.
It means that the resummation scheme passed the consistency check using the gauge-symmetry.


\begin{theacknowledgments}
 This work was supported by the Grant-in-Aid for the Global COE Program "The Next Generation of Physics, Spun from Universality and Emergence" from the Ministry of Education, Culture, Sports, Science and Technology (MEXT) of Japan.
 \end{theacknowledgments}



\bibliographystyle{aipproc}   

\end{document}